\documentclass[reprint, amsmath,amssymb,aps, pra, floatfix]{revtex4-2}
\usepackage[usenames,dvipsnames]{color}
\usepackage{graphicx}
\usepackage{dcolumn}
\usepackage{comment}
\usepackage{bm}

\usepackage[table, usenames,dvipsnames]{xcolor} 
\usepackage{caption}       
\usepackage{colortbl} 
\usepackage{graphicx}
\usepackage{dcolumn}
\begin{document}

\preprint{APS/123-QED}

\title{Question Answering models for information extraction \\ from perovskite materials science literature}

\author{M. Sipilä$^{1}$, F. Mehryary$^{2}$, S. Pyysalo$^{2}$, F. Ginter$^{2}$ and M. Todorović$^{1}$}
\email{milica.todorovic@utu.fi}
\affiliation{$^{1}$Department of Mechanical and Materials Engineering, University of Turku, Finland \\ 
$^{2}$Department of Computing, University of Turku, Finland}

\date{\today}
\begin{abstract}

Scientific text is a promising source of data in materials science, with ongoing research into utilising textual data for materials discovery. In this study, we developed and tested a novel approach to extract material-property relationships from scientific publications using the Question Answering (QA) method. QA performance was evaluated for information extraction of perovskite bandgaps based on a human query. We observed considerable variation in results with five different large language models fine-tuned for the QA task. Best extraction accuracy was achieved with the QA MatSciBERT and F1-scores improved on the current state-of-the-art. QA also outperformed three latest generative large language models on the information extraction task, except the GPT-4 model. This work demonstrates the QA workflow and paves the way towards further applications. The simplicity and versatility of the QA approach all point to its considerable potential for text-driven discoveries in materials research.

\end{abstract}

\maketitle

\section{\label{sec:level1}Introduction}

New sustainable technologies and materials are urgently needed, and materials design plays a key role in their development. In inorganic materials, compositional engineering provides an effective route for designing new materials and tuning functional properties towards intended applications. One group of materials where this is especially utilised is perovskites. They exhibit a promising combination of properties for photovoltaic applications, including  high power conversion efficiencies \cite{snaith2018present, correa2017promises, wu2021main}, which would allow sustainable energy generation \cite{renewable_energy, sustainable_development, bogdanov2019radical}. The stability and functional properties of perovskites critically depend on the material composition. 

The crystal structure of perovskites is characterised by the generic formula ABX$_{3}$, were A and X sites are typically populated by cations and anions and a metal occupies the B site. This allows a very broad range of element substitutions, where organic molecules can also serve as A-site cations (hybrid perovskites). An  entirely new range of functional properties can be accessed via substitutional engineering or alloying, where multiple element substitutions can be implemented on A, B, or X sites. The past decade has seen a proliferation of compositional engineering studies of perovskites to tune their properties, which have resulted in a large number of publications across disciplines and research areas \cite{jeon2015compositional, mbumba2021compositional, laakso2022compositional, bush2018compositional}. There is a need to consolidate known properties of perovskites across different disciplines. This could be achieved by extracting perovskite information from scientific literature with language processing tools.

Natural language processing (NLP) is a field that relies on computation and machine learning to process, generate and analyse human language. In recent years, considerable advancements in NLP have been achieved by transformer neural networks and language models \cite{devlin2018bert}. They exploit transfer learning, where a model is first pre-trained with a large dataset to learn general relationships in text, and then fine-tuned further for a certain specific task. Within NLP, we focus on the common task of information extraction (IE): finding information of interest from non-structured textual sources automatically. 

In materials science, NLP has been used to extract information from publications with named entity recognition (NER) and relation extraction (RE). The aim of NER is to identify entities of interest, such as names or properties of materials from text. With RE, the aim is to determine entities in text, but also their relationships to each other. These tasks can be carried out by supervised machine learning algorithms or rule-based methods. The supervised machine learning NER approach has been deployed to extract materials synthesis parameters \cite{hiszpanski2020nanomaterial}, polymer names \cite{tchoua2019creating}, general materials information \cite{weston2019named}, general solid state materials information, gold nanoparticle synthesis descriptions and doping procedures of materials \cite{matbert} from literature. Similar models have been applied to RE to extract synthesis parameters \cite{kim2017machine}. For rule-based methods, the state-of-the-art is the multi-purpose toolkit ChemDataExtractor2 (CDE2) \cite{chemdataextractor2}, which is capable of RE for materials science. It combines grammar-based parsing rules with the probabilistic Snowball algorithm and was used to create databases about battery materials \cite{huang2020database}, thermoelectric materials \cite{sierepeklis2022thermoelectric} and semiconductor bandgaps \cite{dong} among others.

\begin{figure}[ht]
\includegraphics[width=0.5\textwidth]{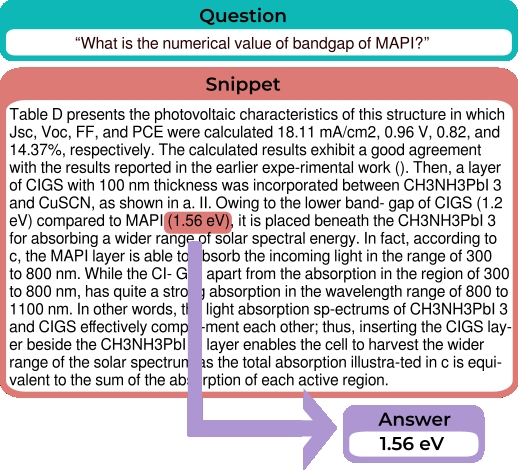}
\caption{\label{fig:snippet} An example of question "What is the numerical value of property of material X?', snippet and the answer, containing a value and units (extracted from \cite{solhtalab2022efficiency}). The name 'CH3NH3PbI3' has been normalised in the snippet to 'MAPI'.}
\end{figure}

Supervised learning methods in NLP require manual annotation of entities to train the models. Analogously, rule-based methods require manual input to construct the syntactic rules by which the information is extracted from text. These present difficulties in applying such methods to different properties and materials: expert knowledge and human effort are needed to retrain the model for each new purpose. Moreover, models tend to focus on processing single sentences \cite{chemdataextractor2, gilligan2023rule}. This leads to loss of information where the relation between entities crosses sentence borders.

Due to their recent advances, generative large language models (LLMs) have also gained popularity in the materials science community. The generative LLMs have been used to extract information about metal-organic framework synthesis \cite{zheng2023chatgpt}, dopants and host materials, metal-organic frameworks and general materials information \cite{dagdelen2024structured} and experimental materials science data \cite{foppiano2024mining}. While generative models offer many possibilities, their main shortcoming is their tendency to produce values or texts not found in the original text. This behaviour, called "hallucination", can lead to incorrect values and unreliable databases. In addition, the most widely used generative models, GPT-3.5 \cite{gpt3.5} and GPT-4 \cite{gpt4} are commercial, so their use in large-scale information extraction would be expensive.

Here we explore the potential of language models inherently incapable of hallucination for IE in materials science. We use language models with the Question Answering (QA) approach. QA is based on the models capability to exploit previous training and identify information that it has not specifically been trained to extract. Starting with a pretrained language model, one can fine-tune it with the general structure of questions and answers to produce a QA model. This tool can then be applied to different domain extraction tasks without need of retraining, in an unsupervised manner, which sets it apart from the task-specific approaches of supervised machine learning or rule-based RE and NER. Trained QA models can return answers to a natural language question (human query) from a context document of arbitrary length, crossing sentence borders. It returns the text span that is most likely the correct answer, otherwise it returns an empty string. While QA models are promising, this technique has not been applied to RE in materials science to date and it is unclear if the QA performance would be satisfactory.

In this study, our objectives were to build a QA framework for extracting information from perovskite materials science literature, evaluate its performance and apply it to a large textual dataset. To implement QA, we considered model choices that range from selecting the question, materials, properties of interest and language models, to deciding what kind of context documents were most appropriate. The context documents are materials science literature segments, referred as snippets throughout this manuscript. Snippets are computationally more efficient to use than full-text publications, which contain a lot of non-relevant text and complex contexts, and could possibly lead to retrieving erroneous information.

The task was to extract material-property relationships as [material, property, value, unit] with the question being 'What is the numerical value of property of material X?'. A typical text snippet with the question and extracted answer is illustrated in Figure \ref{fig:snippet}. For the purposes of testing the model, the property of interest was the bandgap. Bandgap greatly affects the optoelectrical properties of perovskites, which is important for the solar cell research community. To demonstrate the method, we focused on five different perovskite materials: three hybrid (MAPI, MAPB and FAPI) and two inorganic halide (CsPbI$_{3}$ and CsPbBr$_{3}$) perovskites. The narrow focus allowed us keep the initial study compact, evaluate the effect of different model choices and pave the way towards further applications.

A key question was which pre-trained language model to use as the basis of the QA model, since the base model may have a considerable effect on the model performance. We compared five different transformer models, pre-trained with different datasets: base BERT (Bidirectional Encoder Representations from Transformers) \cite{devlin2018bert}, SciBERT \cite{scibert} and three BERT models trained with materials science texts \cite{matbert, materialsbert, matscibert}. Each of the models were fine-tuned towards QA by training with the general domain dataset SQuAD2 \cite{squad2}. This  state-of-the-art QA training dataset contains also empty answers, which allow the model to correctly detect cases where the snippet does not contain any relevant values.  Although SQuAD2 is not targeted towards scientific questions, we relied on the generic capability of the QA models to answer to the questions. The performance of QA systems with different BERTs was compared first to baseline method CDE2 and then to four generative models. 

\begin{figure}[ht]
\includegraphics[width=0.35\textwidth]{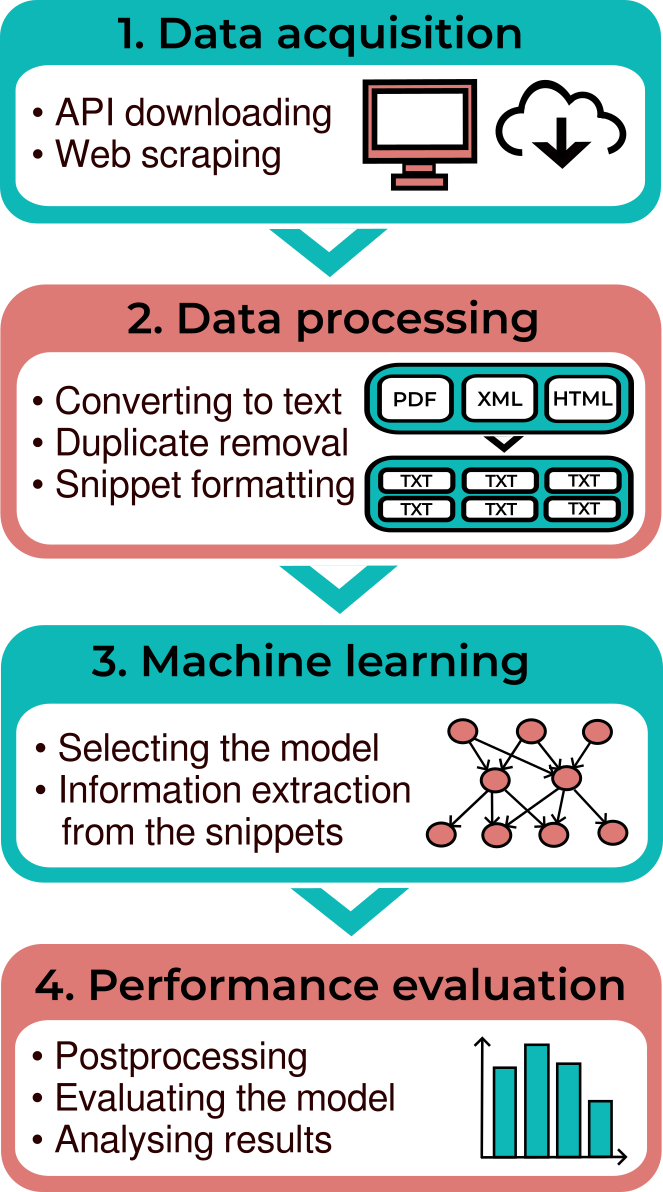}
\caption{\label{fig:workflow}The workflow for implementing and testing the QA method for an IE task.}
\end{figure}

\section{\label{sec:level2}Methods}
Figure \ref{fig:workflow} illustrates the general workflow of our implementation of the QA method for IE. First, the textual data, i.e.\ corpus of scientific publications, was downloaded via publisher application programming interfaces (APIs) and web scraping. Next, this textual data, which was obtained in multiple formats, was converted to plain text and duplicated text was removed. After dividing the text into snippets, we selected a suitable QA model and applied it to the selected snippets. Finally, the QA extracted values were postprocessed, used to evaluate the model and analysed.

\begin{table*}[t]
\caption{\label{tab:table1}%
File formats of publications from different sources, numbers of downloaded texts and conversion tools employed.
}
\begin{ruledtabular}
\begin{tabular}{lcccccc}
\textrm{Source}&
\textrm{Format}&
\textrm{Number}&
\textrm{Conversion tool}\\
\colrule
Elsevier Article Retrieval API & XML & 117,044 & CDE2 ElsevierXmlReader \cite{chemdataextractor2}\\
Springer Nature API & XML & 34,395 & Conversion script\\
Core API & Json & 65,924 & Text selected by json-tag\\
arXiV API & PDF & 3,814 & CDE2 Document package \cite{chemdataextractor2}\\
Royal Society of Chemistry & HTML & 17,254 & CDE2 RscHtmlReader \cite{chemdataextractor2}\\
\end{tabular}
\end{ruledtabular}
\end{table*}

\subsection{\label{sec:level3} Data acquisition}

The first step in IE was to create a dataset of scientific publications. We downloaded publications from 5 sources: arXiV API \cite{arXiV}, Elsevier Article Retriever API \cite{elsevier}, Core API \cite{core, knoth2023core}, Springer Nature API \cite{springer} and also web scraped the Royal Society of Chemistry (RSC) articles \cite{rsc}. Permission to download publications was granted through the university licenses, but permission to scrape publications through the RSC webpage was obtained directly from the RSC. We used the query 'perovskite' or 'perovskites' when downloading and web scraping publications: if a publication matched a word, it was added to the corpus. 

We downloaded a total of 238,431 scientific publications from different sources combined. The number of articles obtained from each source is summarised in Table \ref{tab:table1}. In addition to full texts, we collected the following metadata: year of publication, source and article identifier. In most cases, the article identifier was the digital object identifier (DOI), but 1,161 publications obtained via the arXiV API lacked DOIs, so they were labelled by the \texttt{arXiV id}. Similarly, 12,843 articles from the Core API lacked a DOI and were labelled by the \texttt{Core id}.

\subsection{\label{sec:level4} Data processing into snippet text segments}

Data processing involved three stages: converting data to plain text, removing duplicated publications and formatting snippets.
We converted the scientific publications to plain text because language models cannot process PDFs, HTML or XML tags. We also removed tables and figures from the texts, since these cannot be processed as normal text and were out of scope of this research. There are also currently no satisfactory methods for extracting text from figures, although there are several methods for extracting text from tables \cite{min2024exploring, zhao2023investigating}. Figure captions were preserved. Any supplementary information located in the main manuscript file was included (some Core and arXiV publications). Where the supplementary information was placed in external files, these were omitted from consideration (all other publishers). Converting different publication formats to plain text is a challenging task because the formats of the publications vary, so we used multiple software tools.
Table \ref{tab:table1} describes the original format of publications and summarises the range of conversion tools employed. Here, 89 articles could not be converted with the available tools and were omitted. After conversion, the plain text dataset totalled 195,872 publications.

It is important that the dataset contains only unique publications, to avoid redundancy of extracted values. By using DOI as a unique identifier for publications, we identified and removed 1,505 duplicate manuscripts. 
The articles without DOI were checked against the entire dataset: we converted all plain text articles to numerical vectors of token counts and compared the vector representations to each other.
This allowed us to remove 45 publications without DOI, where the token count vectors were near-identical to another publication. The conversion process and duplicate removal is described in detail in section S1 of the Supplementary Information (SI).
After removing duplicates there were 194,322 unique scientific publications in the dataset.

The plain text articles were segmented into text snippets that contained valuable information. To keep the IE efficient and accurate, the remainder of the manuscript text was discarded at this stage.
First, all the articles bearing this information of value were identified. We used the Elasticsearch search engine \cite{elasticsearch} to query articles by keywords. The keyword list contained the name of the material of interest, the word 'bangap', and their multiple synonyms and acronyms (see Table S2 in SI). From the entire corpus, we retained for further processing only the manuscripts which contained the keywords for material name and property. For example, with the material MAPI we selected 11,193 publications with bandgap information. 
This step is specific to material name and property of interest and was repeated for each material.

The snippet approach was adopted to purge the text of information irrelevant to the task, such as introductory passages, discussions of other materials or references. 
Based on preliminary experiments, we set the length of a snippet to seven sentences. Section S2 in SI describes related information analysis. While most of the information could be found within one sentence (40.4 \% of cases), a substantial portion of snippets had information spanning two (17.8 \%), three (12.7 \%) or more (29.1 \%) sentences. The seven-sentence snippet thus ensures more context than a single sentence and allows to extract also information divided between multiple sentences, although using snippets instead of single sentences somewhat increases the computational requirements of the extraction method. The BERT models have a capability to process texts up 512 tokens in lenght without loss of information, and 7 sentence snippets usually fall below this limit. We carried out tests to confirm this (details in SI table S4).

To build the snippets, we had to ensure they contained at least one mention of the material name (for example, 'MAPI'), the property name ('bandgap') and the unit name ('eV').
Because the relevant terminology varies, we normalised the text by standardising pertinent nouns to one chosen form. Here, the previously-constructed list of synonyms was used to convert all materials and property names to e.g. 'MAPI' and 'bandgap'. For each article in the corpus subset, we extracted seven-sentence snippets which contained mention of material, property and unit. The snippets were extracted by moving a seven-sentence text window down the manuscript text and saving the snippets which contained the necessary information (see SI section S2). For example, for MAPI material we extracted 7,281 snippets from the 11,193 articles parsed.

\begin{figure}[hb]
\includegraphics[width=0.5\textwidth]{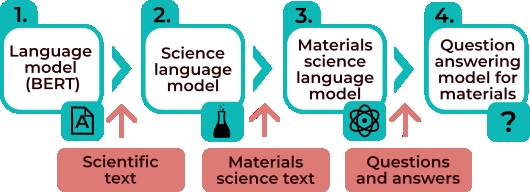}
\caption{\label{fig:transfer_learning} A schematic figure of transfer learning that underpins the QA model for materials science.}
\end{figure}

\subsection{\label{sec:level5} QA model implementation}
The state-of-the-art methodology in the field of NLP are transformer neural networks based on self-attention \cite{vaswani2017attention}. This enables the language model to learn relationships between words and the context of the word, regardless of the text length. In this study, we opted for the encoder-only transformer model, since it is intended for analysing textual data (decoder-only and encoder-decoder models are designed for text generation). A major advantage of encoder-based models in IE is their inability to hallucinate answers, which makes them a valid choice in tasks where the answer is directly a substring of the snippet \cite{devlin2018bert}.

Transformers rely on transfer learning to acquire new model knowledge and adapt to more specific tasks. This approach allows models, which learned useful features, contexts or patterns from a previous dataset, to retain previous information and add to it with the further training. Figure \ref{fig:transfer_learning} illustrates the process of training a base language model in stages with a carefully selected set of domain-specific texts. It is common to start with the generic encoder-only language model BERT, trained on general-domain English language texts, such as English Wikipedia, books and text scraped from multiple web sources \cite{devlin2018bert}. This model can be further trained with scientific texts to generate a science English language model, which can then be trained with materials science texts. 
In our approach, the final language model was further trained for question answering tasks to produce the QA model. 

Different base models may have a profound impact on the performance of the QA model. Since there is no literature precedent, we tested five different language models: two general ones and three materials science BERTs. We tested the base BERT model, trained with general-domain English language text \cite{devlin2018bert}, and the scientific language model SciBERT, base BERT trained with scientific text from the computer science and biomedical domains \cite{scibert}. The materials science models were MatBERT \cite{matbert}, MatSciBERT \cite{matscibert} and MaterialsBERT \cite{materialsbert}. MatBERT was trained from base BERT with randomly sampled two million documents, mostly consisting of peer-reviewed materials scientific journal articles  \cite{matbert}. On the other hand, MatSciBERT was trained from SciBERT \cite{scibert} with 153,978 scientific publications about inorganic materials. MaterialsBERT was trained based on PubMedBERT \cite{pubmedbert} with 2.4 million scientific abstracts from the multiple materials science sub-domains. 

Following the standard practice in application of encoder models to span retrieval tasks, we fine-tuned the above language models for the QA task with the SQuAD2 \cite{squad2}. During fine-tuning, the BERT-based model is assigned two probabilities to each token in the context document, indicating the likelihood of it being the start or end of the answer. 
The text span between the two tokens that are most likely to start and end the answer is returned as the QA answer. \cite{devlin2018bert}. If there is no answer, the beginning and ending token as predicted to the the [CLS] token, which is prepended to all inputs in the model. Computational implementation was based on the Huggingface transformers-library \cite{transformers} version 4.20.1 and Pytorch version 1.12 \cite{article_code}. All computing was performed on the NVIDIA Volta V100 GPU. The SQuAD2 features 107,785 question-answer pairs, as well as 53,775 unanswerable questions, on 526 Wikipedia articles. Section S3 in SI summarises the hyperparameter \textcolor{black}{tuning} procedure.
In applications, we accounted for statistics by using 5 different seeds for weight initialisation in the training process to build and test 5 QA models for each BERT.
The average values and standard deviation were reported as the final performance metrics. 
The trained QA model, when provided with a question, generates a number of answers with a model confidence score associated with each one. 
By default only the top answer is retrieved since it is most likely to be correct (as illustrated in Figure \ref{fig:snippet}), but the workflow can be adjusted to retrieve multiple answers. Should the correct value be missing in the text, we would expect an empty string to be returned, based on SQuAD2 training.

\subsection{\label{sec:level6} Performance evaluation}

Before applying the QA model to the complete dataset, we benchmarked the performance of the different language models in the QA framework to identify the \textcolor{black}{best performing} BERT. For this, it was necessary to postprocess model answers, evaluate suitable peformance metrics and visualise the retrieved data.

In applied NLP, answers are typically postprocessed to standardise them and facilitate the computation of performance metrics. In response to our query prompt 'What is the numerical value of bandgap of material X?', the model could return a number of answers.  We expected the correct answer to contain the correct numerical value of the bandgap and the corresponding unit 'eV', but text answers, numerical ranges and conjunction words were also possible and valid. Several postprocessing schemes were devised and tested to standardise correct results (see section S4 of the SI for details). For minimal processing and optimal outcomes, we  
omitted from consideration all the answers without any digits and with letters outside the set 'e, v, t, o, a, n, d'. The rest of the results were evaluated against gold standard answers to compute metrics. After this, any values denoting a range were averaged (extracted range '1.5-1.6' eV was recorded as 1.55 eV) to ensure that the extracted values would not carry bias towards the ends of the range simply because of rounding practices.

Human opinion sets the gold standard in NLP tasks. From the complete dataset, we  manually annotated a subset to establish the correct answers. The 600 text snippets were selected to evenly balance bandgap information on the five materials considered. Six materials science experts performed the annotation (including empty answers) so that each snippet was reviewed by two persons. The process is described in section S5 of the SI. After agreement was achieved, 175 snippets were found to contain 209 numerical bandgap values, and 425 featured none. Out of 175 snippets, the annotators identified 151 snippets with a single bandgap value, and 24 snippets with two or more (up to 6) values.  

The annotated dataset was used to compute classification accuracy metrics for the QA models. A true positive was achieved if the QA result was numerically equivalent to the manually annotated value. This way the QA returned answer '1.5' would be considered a true positive for an annotated value of '1.5 eV', but a returned value of '5' would not. Similarly, a true negative denoted a correctly returned empty text span. These metrics were employed to compute precision (the fraction of correct answers from all retrieved answers), and recall (the fraction of correct returned answers compared to all golden standard answers). We used F1-score, the harmonic mean between precision and recall, to identify the best QA BERT model. Results were compared against the well-established CDE2 code (version 2.1.2) for IE, which was applied to the annotated dataset following previously established procedures \cite{dong}. 

Next, we compared the QA results to the results obtained with four different generative models: Mixtral-8x7B-Instruct-v0.1 \cite{jiang2024mixtral} (Mixtral-8), Llama3-ChatQA-1.5-8B \cite{liu2024chatqa} (Llama3), GPT-3.5 Turbo (GPT-3.5) \cite{gpt3.5} and GPT-4-0613 (GPT-4) \cite{gpt4}. Mixtral-8 is based on the language model Mixtral-8x7B, which has been demostrated to match or outperform GPT-3.5 and trained to follow instructions \cite{jiang2024mixtral, mixtral-instruct}. Llama3-ChatQA-1.5-8B is trained from Llama-3 base model (second latest Llama model) to excel at conversational question answering \cite{liu2024chatqa}. The results were generated using zero-shot prompting, optimised over four different prompts and postprocessed (see SI S7). Results were extracted using temperature 0 to ensure the highest possible determinism of the models. Our prompt engineering tests revealed that different models require different prompts for the optimal results. Moreover, we established that even with temperature 0 and fixed seed (12), the GPT models exhibited different results for repeated calculations \footnote{This is an unexpected, nevertheless publicly documented feature of the OpenAI API. \textit{platform.openai.com/docs/advanced-usage/reproducible-outputs} and the seed parameter is still in Beta phase \textit{platform.openai.com/docs/api-reference/chat/create}}. Consequently, for GPT computations we performed 4 repetitions with identical settings and prompts, and  we report the average and the standard deviation of the answers.

Since the QA model returns the text span with the highest confidence score (the most likely answer), evaluating performance on snippets with a single bandgap value was straightforward. However, it was unclear how to approach the 24 snippets with up to 6 bandgap values. An easy solution would have been to acknowledge any correct value returned, but this would not have allowed us to extract all possible answers, which is desirable in applications. For a more complete outcome, we shifted from considering just one result with the highest confidence score to considering several. From the many prospective answers returned by the model, we retrieved 6 with the highest confidence scores to ensure that all bandgaps could be extracted from each snippet in the annotated dataset. 

At this stage, we introduced the QA confidence threshold as a model parameter. A closer examination of the confidence scores associated with the snippet answers revealed that model certainty varies considerably for different snippets, and this should be taken into account (because less confidence points to a likely wrong answer). We examined model performance with different confidence thresholds applied to the top 6 answers: 0.0125, 0.025, 0.05, 0.1, 0.2, 0.4 and 0.8. 

Here, all of answers with the confidence score above the threshold were retained and performance metrics were computed (up to a maximum of 6), while the rest were discarded. We performed a 4-fold cross-validation (CV) using a manually annotated dataset of 600 snippets to determine the optimal threshold for each QA model. The dataset was split into four folds, with three folds used as the validation set in each iteration, and the remaining fold serving as the test set. The folds were shuffled between each iteration so that each fold served once as a test set. The final results were averaged across all iterations.
Low thresholds would permit most answers at the expense of accuracy, which increases recall but lowers precision. The opposite is true for high thresholds, so considering the entire range is a helpful test to establish which confidence threshold best balances precision with recall, optimising F1-scores. For each cross-validated language model, we identified the optimal threshold on the CV validation set, then applied it to the CV test set to compute the final evaluation metrics.

All extracted answers comprised a distribution of bandgap values for a particular material. We performed a statistical analysis for the range, mean and median values. When extracting bandgap values with CDE2, only values corresponding to the set of five perovskites were selected, in order to realiably compare the QA method to CDE2.

\section{\label{sec:level7}Results}

\begin{figure*}[ht]
\includegraphics[width=\textwidth]{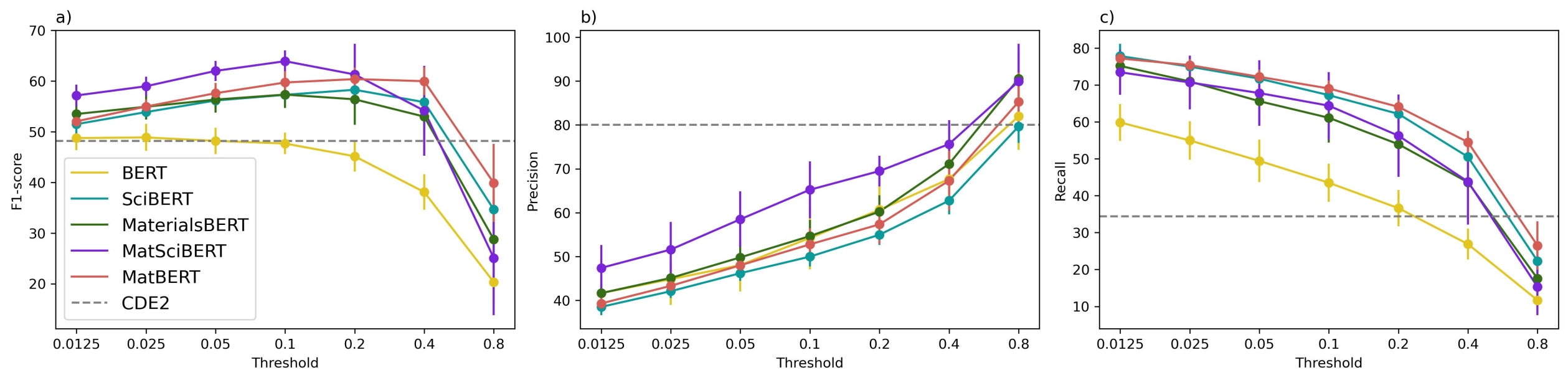}
\caption{\label{fig:thresholds} The evaluation metrics of the \textcolor{black}{CV validation set} answers returned by different language models trained towards QA task by confidence score threshold: (\textbf{a}) F1-score, (\textbf{b}) precision and (\textbf{c}) recall. The grey dashed line denotes the CDE2 performance with a \textcolor{black}{48.2} F1-score, precision at \textcolor{black}{80.0} and recall at \textcolor{black}{34.5}. }
\end{figure*}

The predictive power of QA for bandgap extraction was tested with respect to model choices, namely different BERTs. 
We first address the model quality tests with different QA confidence thresholds. Performance metrics (precision, recall and F1-score) were evaluated on the annotated dataset for all five BERTs and compared to CDE2 and four generative model values. Next, all QA models were applied to limited-scale IE to examine the effect of model choices on the range the extracted values. Based on tests results, we selected optimal QA model and applied to the complete corpus to demonstrate QA-based IE. 

\subsection{\label{sec:level8} Evaluating QA model performance}

The predictive power of the QA models on the CV validation dataset is illustrated in Figure \ref{fig:thresholds}. Precision, recall and F1-score were computed for all five BERTs as a function of confidence threshold, and compared to the CDE2 results. The values in Figure \ref{fig:thresholds} with their standard deviations are in SI Table S8. When the threshold increases, model precision rises, but recall decreases. This is expected behaviour, because high thresholds enforce model certainty, and simultaneously reduces the number of answers returned. Overall, the general BERT achieves the poorest outcomes due to fewest answers returned (inferior recall). 

The optimal F1-score of 63.9 was observed with QA MatSciBERT (at threshold 0.1), which can be associated with consistently high precision of this language model (Figure \ref{fig:thresholds}b). 
In contrast, MatBERT produced consistently high\textcolor{black}{est} recall (Figure \ref{fig:thresholds}c). MaterialsBERT was the least effective of the three materials language models and performed similarly to SciBERT on this task. We note that CDE2 favours precision over recall (by construction) and thus did not reach F1-scores beyond 48.2. QA models outperformed CDE2 on precision only at very high tresholds, but at the cost of much degraded recall and F1-scores. Throughout the different thresholds, MatSciBERT exhibited high standard deviations, which is evident from the data uncertainties in Figure \ref{fig:thresholds} b and c. Elsewhere the standard deviations are smaller, pointing to a reduced dependence on initial data in model training.

In Table \ref{tab:testset}, we observe that different confidence thresholds optimise the QA extraction for different BERTs. The top F1-score 63.9 obtained with MatSciBERT and threshold 0.1 is closely followed by F1-score of 60.2 for MatBERT (threshold 0.2). The F1-score range of 57-58 indicates a similar performance of BERT, SciBERT and MaterialsBERT. All language models except BERT achieved best F1-scores at similar values, 0.1 or 0.2 in confidence. This finding was clarified once we computed the average confidence scores of the top answers (Table S10 in SI). While the average first answer confidence score was 0.7, the average second answer confidence score was as low as 0.06. It follows that that simply selecting the one top answer could produce good extraction, without the need to consider confidence score thresholds. To test this, we calculated the evaluation metrics when selecting only the one top answer from the QA results (see Table S9 in the SI). The best F1-score obtained with this approach was 62.8 with MatSciBERT. The results indicate that selecting just one top answer (the recommended QA approach) produces F1-scores that are very similar to the best threshold results. The only exception was the BERT QA model, where the low confidence score threshold produced higher F1-score.

\begin{table}[ht]
\centering
\caption{\label{tab:testset}
Evaluation metrics F1-score (F1), precision (P) and recall (R) of test set for each QA model with the best validation set threshold (T). MaterialsB. denotes MaterialsBERT and MatSciB. MatSciBERT.}
\begin{ruledtabular}
\begin{tabular}{lcccc}
\textrm{}& \textrm{T} &\textrm{F1}& \textrm{P}& \textrm{R}\\ \hline
BERT & 0.025 & 57.2 (± 2.2) & 47.4 (± 5.2)  & 73.5 (± 6.1) \\
SciBERT &  0.2 & 58.1 (± 3.7) & 54.9 (± 4.3) & 62.0 (± 4.5)\\
MatBERT & 0.2 & 60.2 (± 4.1) & 57.2 (± 5.7) & 63.9 (± 4.5) \\
MaterialsB. & 0.1 & 57.3 (± 4.0) & 54.9 (± 5.4) & 60.9 (± 7.8) \\
MatSciB. & 0.1 & 63.9 (± 3.5) &  65.4 (± 8.0) & 64.4 (± 9.3)\\
\end{tabular}
\end{ruledtabular}
\end{table}
\vspace{1cm}
\arrayrulecolor{black}

\begin{figure*}[ht]
\includegraphics[width=1\textwidth]{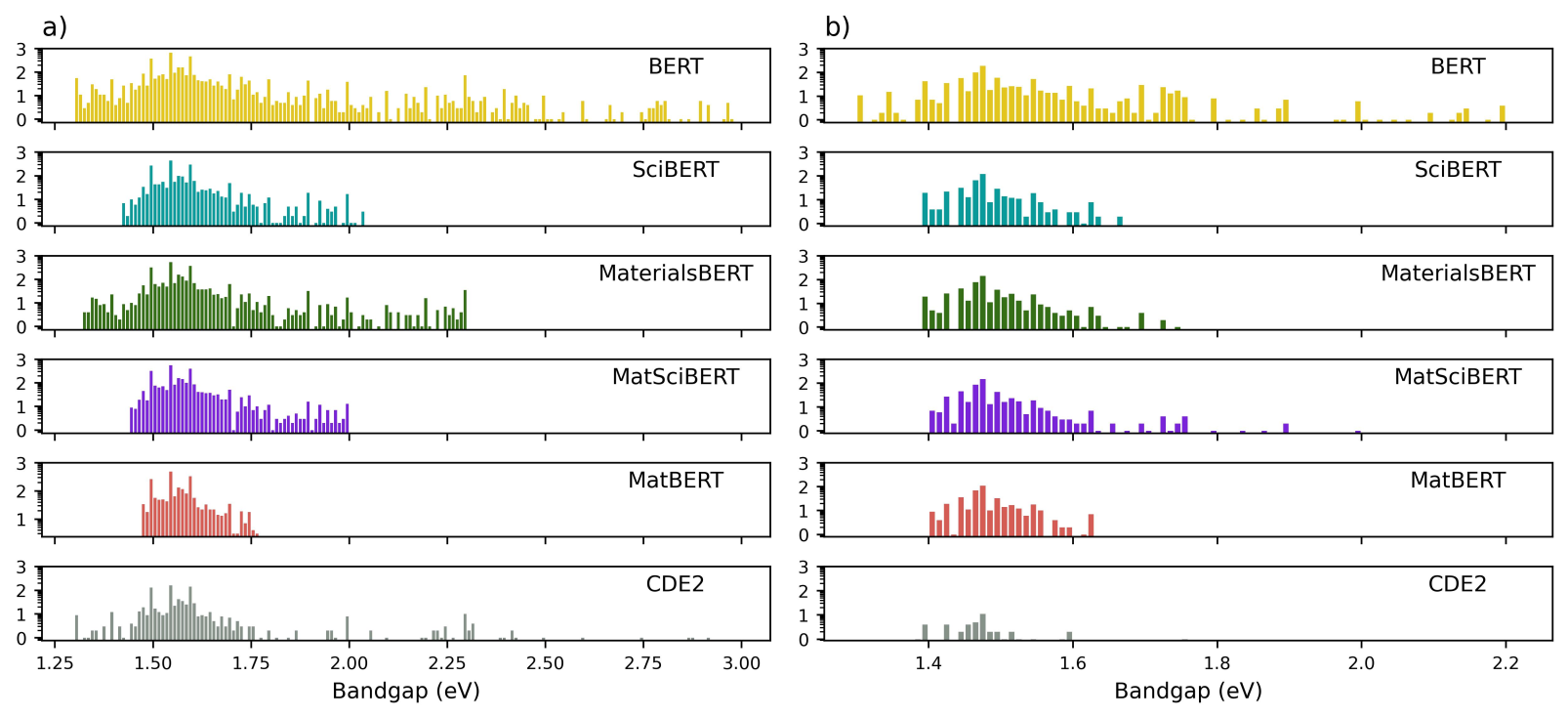}
\caption{\label{fig:flat_histograms}Histograms of extracted bandgap values with QA models based on BERT, SciBERT, MaterialsBERT, MatSciBERT and MatBERT for a) MAPI and b) FAPI hybrid perovskites. The y axis was modified by  $\mathrm{log}_{10}$ for clearer comparison. The CDE2 values have been postprocessed using the CDE2 postprocessing script in previous work \cite{dong}. The range of x-axis was set to match the broad value range of BERT QA answers, which nevertheless excluded several CDE2 values from view (97 for MAPI and 5 for FAPI)}

\end{figure*}

The performance metrics indicated differences between the BERTs tested on the annotated dataset (600 snippets). Still, it is unclear to what extent this influences IE in real-world applications, with thousands of context documents. To explore the scale of this effect, we applied the 5 QA language models and CDE2 to our full dataset, and extracted the values based on the best thresholds defined in Table \ref{tab:testset}. We focused on the two materials with the most and least snippets available: MAPI had extensive literature coverage (7,283 snippets) while FAPI was least featured (1,251 snippets). Here we expect both precision and recall to play a role in the number of extracted values and their distribution. On average, it required 2.7 h and 0.5 h to extract MAPI and FAPI values respectively on a single NVIDIA Volta V100 GPU with QA models, and 3.9 h and 0.7 h with CDE2. 

The bandgap distributions presented in Figure \ref{fig:flat_histograms} were subject to statistical analysis for the number of extracted values (Table S12 in the SI). In the case of MAPI, all models recorded the mode of 1.55 eV and median of 1.56 or 1.57 eV. Among QA models, base BERT bandgaps were most diverse with a range of 1.70 eV, while MatBERT returned a narrow range of 0.28 eV. A broad range suggests many retrived values: QA BERT extracted 3,995 MAPI bandgap values. For FAPI, we observed similar trends: QA BERT obtained the maximum of 993 bandgap results in contrast to QA MatBERT with 397 values. This was reflected in their ranges of 0.93 eV and 0.22 eV respectively. The FAPI mode and median were 1.48 eV for all the models except BERT, where the median 1.49 eV was affected by the large range of values.

While a large number of retrieved values is always desirable, an overly broad range of extracted bandgaps may not indicate quality. Large bandgaps over 2 eV identified for MAPI by BERT and MaterialsBERT are unlikely and indicate lack of precision in extraction. MatSciBERT and MatBERT exhibited the best performance in this test.  Here, we additionally clarified if the extremes of the distribution ranges contained correct bandgap values from text, or erroneous answers. We manually selected 10 random QA MatBERT retrieved MAPI bandgaps from the high range (above 1.7 eV) and low range (below 1.49 eV) and inspected the original snippets. In all 10 examples the extracted values were correct bandgaps, but in two cases MAPI had been doped by bromine, which resulted in a bandgap shift. The latter observation made clear that many scientific factors may produce a genuine difference in bandgaps reported in literature.

Based on the performance evaluation tests, we opted for the MatSciBERT in all subsequent IE tasks. 
It exhibited the highest F1-score in the Table \ref{tab:testset}. Based on Figure \ref{fig:flat_histograms} and SI Table S9, MatSciBERT extracted much information (2,733 MAPI bandgaps) from the second most narrow numerical range, allowing less scope for error.

To evaluate how QA compares with state-of-the-art tools, we applied four generative models to the same information extraction task with the whole evaluation set (600 snippets). The metrics are summarised in Table \ref{tab:gen_results} and indicate that F1-scores achieved by generative models varied greatly. The optimal F1-score of 69.8 was obtained with the GPT-4, improving upon GPT-3.5 and surpassing also the QA MatSciBERT results. In contrast, two popular high-quality models, Mixtral-8 and Llama3 exhibited lower F1-scores on account of poor precision. This observation is explained by the high number of hallucinated answers for Llama3 (161) and Mixtral (39), which are considered as false positives and lower precision. Generally, the recall of generative models is similar or better to the QA models. Data suggests that only the paid GPT-4 model surpasses the QA MatSciBERT on information extraction tasks. Additionally, we inspected the only hallucinated value from GPT-4 and found it to be "2.39 eV". It originates from the sentence "Above bandgap photoexcitation was performed using a Duetto laser at a photon energy of 3.49 eV (i.e. 1.1 eV above the direct bandgap excitation)" \cite{cannelli2021quantifying}, where it appears that GPT-4 had calculated the bandgap value to be 3.49 eV - 1.1 eV. So it seems that the value was correctly calculated even though it was not directly stated in the article.

\begin{table}[htbp]
\centering 
\caption{\label{tab:gen_results}
F1-score (F1), precision (P), recall (R) and number of hallucinated answers (H) from the generative models, CDE2 and the best QA model MatSciBERT (QA).}
\begin{ruledtabular}
\begin{tabular}{lcccc}
\textrm{}& \textrm{F1}& \textrm{P}& \textrm{R}& \textrm{H}\\ \hline
Mixtral-8 & 54.1 & 43.6  & 71.3 & 39\\
Llama3 &  42.2 & 30.5 & 68.4 & 161\\
GPT-3.5 & 53.1 (± 0.3) & 40.9 (± 0.6) & 76.0 (± 2.3) & 21.3 (± 12.6)\\
GPT-4 & 69.8 (± 0.6) & 61.4 (± 0.9) & 80.9 (± 0.0) & 1.0 (± 0.0)\\
QA & 63.9 (± 3.5) & 65.4 (± 8.0) & 64.4 (± 9.3) & 0\\
CDE2 & 48.2 & 80.0 & 34.5 & 0\\
\hline
\end{tabular}
\end{ruledtabular}
\end{table}
\vspace{1cm}

\begin{figure*}[ht]
\includegraphics[width=\textwidth]{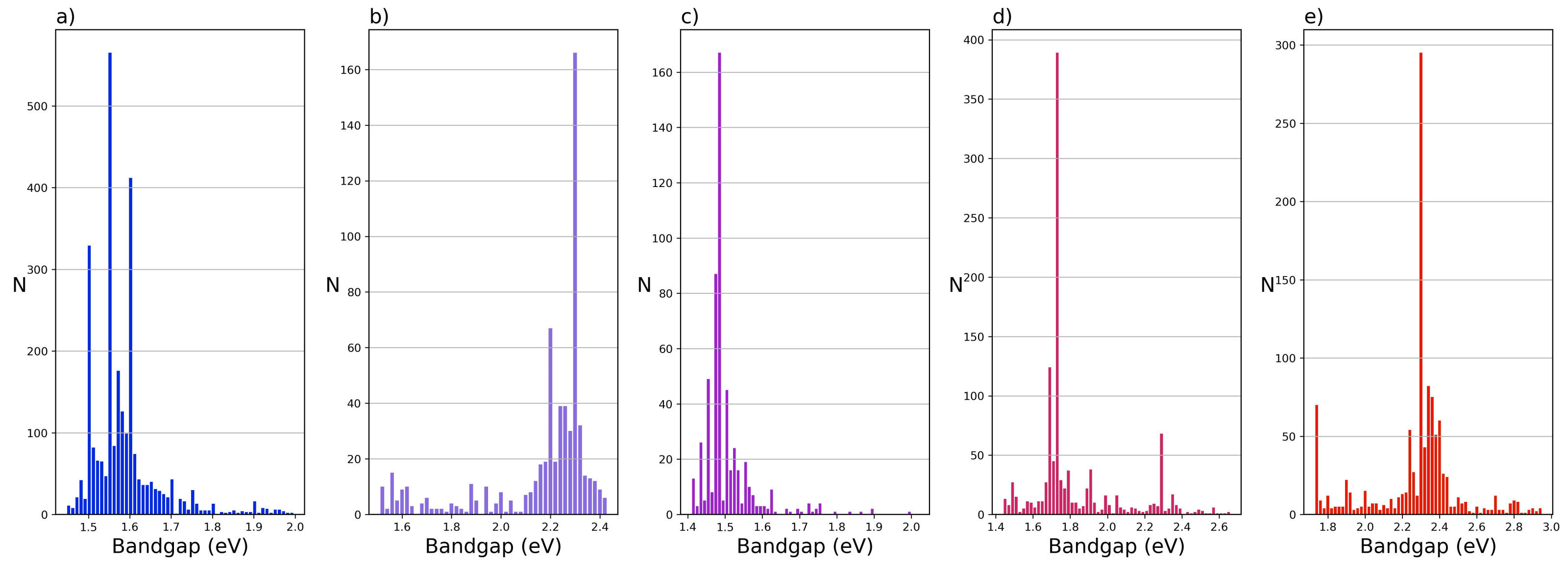}
\caption{\label{fig:histograms} Histograms of bandgap values extracted with MatSciBERT: a) MAPI, b) MAPB, c) FAPI, d) CsPbI$_{3}$ and e) CsPbBr$_{3}$.}
\end{figure*}

\subsection{\label{sec:level9}Information extraction of perovskite bandgap values from the full dataset}

\begin{table}[b]
\caption{\label{tab:statistics}%
Number of extracted snippets (Ns), MatSciBERT extracted values (Nv), mode of extracted values after postprocessing and literature references for comparing the extracted values.
}
\begin{ruledtabular}
\begin{tabular}{lcccc}
\textrm{Material}&
\textrm{Ns}&
\textrm{Nv}&
\textrm{Mode (eV)}&
\textrm{Ref.}\\
\colrule
MAPI & 7,283 & 2,733 & 1.55 & \cite{belayachi2021study, weber2016phase, mohan2024static}\\
MAPB & 1,646 & 641 & 2.30 & \cite{rehman2015charge, qiu2020perovskite, adonin2018antimony}\\
FAPI & 1,251 & 536 & 1.48 & \cite{masi2020chemi, eperon2014formamidinium, yang2016effects}\\
CsPbI$_{3}$ & 1,734 & 1,128 & 1.73 & \cite{ahmad2017inorganic, ye2020stabilizing, parida2019two}\\
CsPbBr$_{3}$ & 2,029 & 1,125 & 2.30 & \cite{yuan2018enhanced, liu2019growing, liang2016all} \\
\end{tabular}
\end{ruledtabular}
\end{table}

We applied QA MatSciBERT to extract bandgap values for five perovskite materials: three hybrid and two inorganic ones. The number of snippets (Ns) and number of extracted values (Nv) are presented in the Table \ref{tab:statistics}. The resulting distributions are depicted in histograms in Figure \ref{fig:histograms}. 
We first verified the overall accuracy of extracted data. The mode of the extracted values agrees well with the experimental values found from the literature sources (Table \ref{tab:statistics}). 
MAPI data featured higher peaks at 1.5 and 1.6 eV in Figure \ref{fig:histograms}a) (MAPI), which can be explained by the trend of researchers reporting values in two significant digits. 

\begin{figure}[ht]
\includegraphics[width=0.5\textwidth]{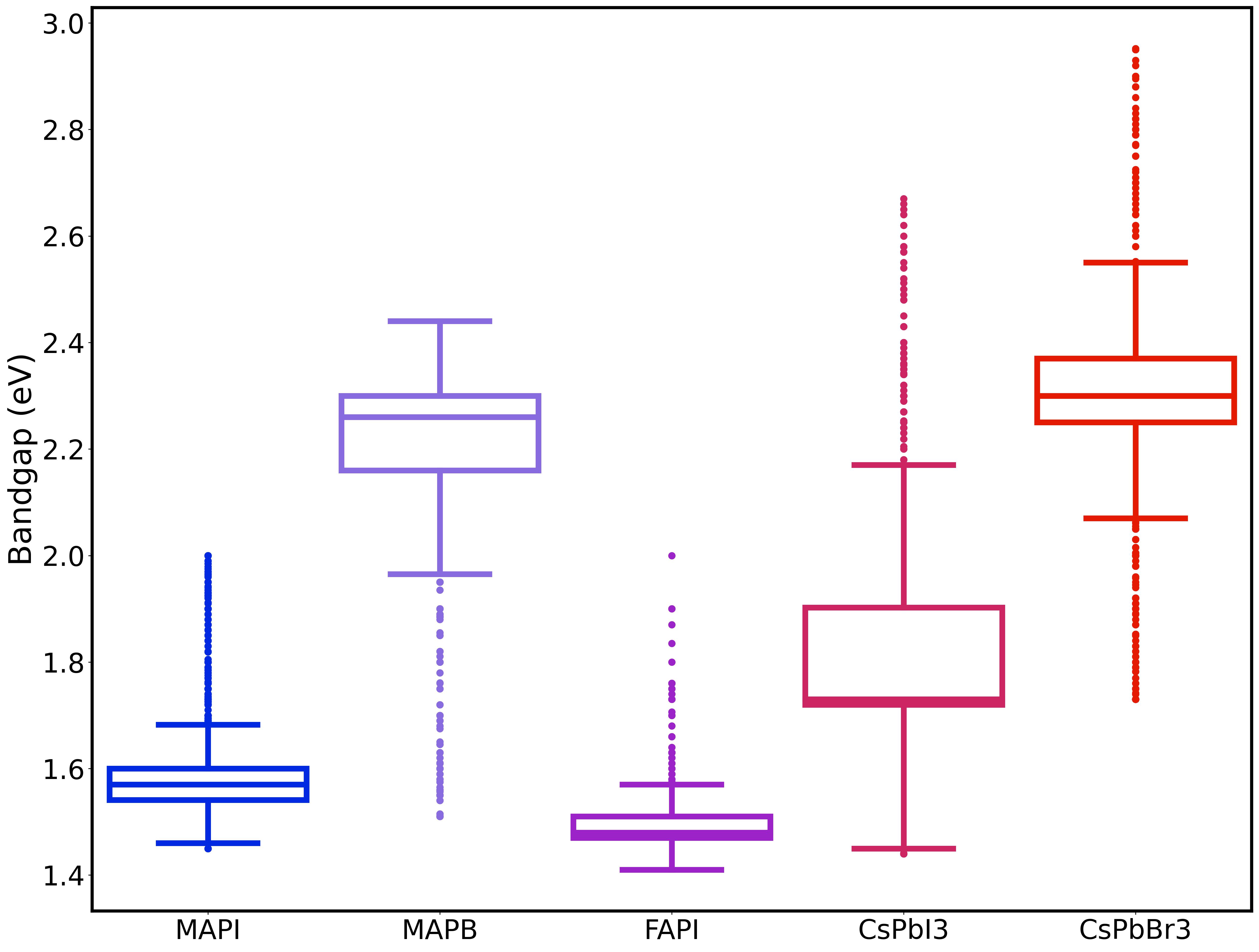}
\caption{\label{fig:boxplot} Comparison of the bandgap distributions extracted for different perovskite materials. \textcolor{black}{The line inside the box denotes the median of the distribution. It divides the box into the upper quartile 25 \% (upper half), and lower quartile 25 \% (lower half).The whiskers mark the remaining two 25 \% quartiles and the dots indicate outliers.}}
\end{figure}

The statistical bandgap distributions for different materials are compared in the Figure \ref{fig:boxplot}. Statistical analysis can be found in Table S13 of the SI. Most of the observed distributions featured a broad spread and data bias. Bias towards lower values and lack of low outliers was observed for the hybrid perovskites (MAPI and FAPI). The behaviour for inorganic perovskites was less clear: CsPbI$_{3}$ and CsPbBr$_{3}$ bandgap data presented long tails towards high bandgap values.  

The ranges of extracted bandgap values could vary considerably. Overall, the ranges for hybrid perovskite bangaps were smaller than those for the inorganics. We compared distribution spreads to the number of extracted values in Table \ref{tab:statistics} to establish if more values simply leads to broader distributions. However, this was not the case, with MAPI registering both a large number of extracted data and a narrow spread. As previously noted, there could be many factors behind the spread of extracted values, and this may not be a concern in database applications as long as the averages values are accurate.
 
\section{\label{sec:level10}Discussion}

We had set out to explore the potential of QA models for IE in materials science, given the task to extract perovskite bandgaps from literature. We found the QA models to be straightforward to implement, which is an advantage. The only model training required was with the general QA dataset, which was easy and fast. They are also versatile, and could be directed towards specific knowledge domains via different base BERTs.  

The different tests demonstrated that QA models are capable of IE from materials texts with good accuracy. Optimising confidence score thresholds with CV produced higher accuracies than just extracting one answer from one snippet. This finding likely arises from multiple bandgap values found in several single snippets.
Varying the threshold would also allow to tune QA models to emphasise either precision or recall, depending on the task at hand. In addition to threshold optimisation, we calculated the F1-scores of different models for top first answer. The relatively small differences between these results and the optimised threshold results can be explained by the small fraction of the snippets with more than one bandgap value in our evaluation dataset.

The choice of the QA language model was most important for the accuracy of information retrieval. In our tests, base BERT was least optimal, most likely because it had not been trained with domain literature in materials science. We expected the three materials BERTs to perform the best, but here we also observed differences. The F1-scores of QA MaterialsBERT were slightly lower and it extracted a broader range of numerical values in application tests, which could lead to erroneous answers. QA MatBERT values exhibited the shortest range in the application test, but its F1-score was not as high as that of QA MatSciBERT, which also performed well on the extraction test.

The differences between QA models can be explained by the different training sets of the base language models. MatBERT was trained with 2 million publications from a broad materials science domain, which is much more text than the 2.4 million materials science abstracts used to train the MaterialsBERT. Also, abstracts can lack some concepts or words which full text articles contain. MatSciBERT training set was smaller (153,978 publications), but the underlying SciBERT may have produced the high F1-score of the model by contributing the capability to process scientific text. Further tests are needed to establish if the comparative BERT performance carries over to different QA extraction tasks and datasets. 

It was surprising to discover that the F1-scores of all five BERT language models exceeded the CDE2 baseline. We note, however, that CDE2 was optimised towards a different IE task: to retrieve all materials properties for all materials found in texts. This may render it less sensitive to a particular property and material, a task made specific by the textual query we used to guide QA extraction. This also limits the QA procedure: the user must identify the material of interest in advance, as the algorithm does not extract values for all materials in the snippets like CDE2 does. The performance gap could be due to another practical difference: although CDE2 operated on entire snippets, it was designed to extract information from single sentences only. Text search analysis revealed that all necessary information was contained in one sentence for only 40.4 \% of the snippets. That might explain lower overall F1-scores observed for CDE2. In precision of extracted values, CDE2 still outperformed the QA models.

\textcolor{black}{There are substantial differences in the performance of generative models, as evident from Table \ref{tab:gen_results}. Unlike the models used through OpenAI generation API, Mixtral-8 and Llama3 models allow full control over the result generation and exhibit deterministic behaviour, but the F1-score of Llama3 falls short. Deterministic behaviour is desirable in IE tasks for reproducibility purposes. 
In terms of F1-score, the GPT-3.5 model performs similarly to the QA models, but could not reproduce answers. The GPT-4 model F1-score exceeded the QA MatSciBERT F1-score at 69.8 to 63.9, which highlights the power of the one of the largest generative models. However, as long as the GPT-4 model is not open access and requires payment, the gain in terms of F1 would need to be weighted against the costs for very large-scale runs, where the comparatively light-weight QA model might have an advantage. The tendency to hallucinate answers is more evident with Mixtral-8 and Llama3, but even a small fraction of hallucinated answers (GPT-3.5 and GPT-4) is undesirable. Additional prompt engineering could be deployed to further reduce hallucinations, but this remains out of scope of our study. } 

An advantage of the QA method is that it can be adapted to different extraction tasks. The QA models were not explicitly trained to extract the value of 'bandgap', but were capable of doing so with relatively high accuracy. The QA models could now be used to retrieve also other properties without need to retrain the transformers. This presents an attractive alternative to more established machine learning methods for IE, where a large amount of manually annotated text is typically required to retrain models for each new IE task. The QA framework can effectively be used as an unsupervised machine learning tool. From the end user point-of-view, QA models require only the question prompt and the context documents, which makes them more accessible to non-expert users. 

Information extraction of materials properties inevitably results in a numerical distribution. With unsupervised use of QA, it is important to critically evaluate the results. 
The extracted values might deviate from the median for multiple reasons: experimental measurement conditions may vary, old references may contain incorrect measurement values, and  computational results differ from the experimental ones. Previous work \cite{vona2022electronic} has established that the computational bandgap of PbI$_{2}$ varies from 1.4 eV to nearly 3.6 eV depending on the computational approach. Since PbI$_{2}$ forms part of MAPI and CsPbI3, the computational bandgap results might vary considerably. A broad distribution does not necessarily mean poor performance. Still, if the range is very wide, there is no means to differentiate the origin of the values without supervision, so it is recommended to favour QA models that produce a narrower range. While our demonstrated application was limited in scope to five materials, it indicates the predictive power of the QA approach and will serve to guide further extraction work.

QA models seem to derive additional performance from the context found in surrounding texts. This could be attributed to the latent knowledge conferred by their base models. While this is an advantage, working with snippets can also present an application bottleneck. In this study, snippet generation was specific to material name and property of interest (bandgap). This step was designed to remove any irrelevant textual information, consequently a search for any other materials property with the same snippet corpus would have been inappropriate (and not as accurate). Applying QA models to other extraction tasks may not require retraining the models, but it would require generating new snippets from the full corpus. In future studies, snippet generation with the same Elastic search approach could be automated to require minimal human involvement. Additionally, several aspects of the workflow could be parallelised for greater computational efficiency.

\section{\label{sec:level11}Conclusions}
 
We constructed a workflow to extract bandgaps from materials science articles with the QA method, where a language model extracts information from the context document based on the question asked. We collected a large textual dataset of scientific publications on perovskites and trained five different BERTs to perform the QA task. Models were compared on the task of extracting materials bandgaps from the scientific literature. The MatSciBERT-based model outperformed BERT-, SciBERT-, MaterialsBERT- and MatBERT-based models on the manually annotated evaluation dataset, exceeding the performance of CDE2 and three of the state-of-the-art generative models. GPT-4 exhibited slightly better performance than QA (69.8 and 63.9 respectively), but also a lack of reproducibility, and additionally required payment and prompt engineering. The MatSciBERT QA model was applied to bandgap extraction from a database of 194,322 scientific articles for five prototypical perovskites. The method was easy to apply and results were accurate.

This study demonstrates the suitability of the QA method for extracting material-property relationships from materials science literature. 
QA-based IE was accurate and efficient with materials science literature snippets and reached F1-scores above the state-of-the-art. 
Moreover, QA allows extraction by human language queries, which would lower the barriers for non-expert users and promote diverse applications. Further tests are needed to explore the performance of QA tools on different tasks. The versatility of the approach indicates  considerable potential of QA for NLP applications in the field of materials science, which could be deployed to accelerate discovery and materials design. 
\\

\section{\label{sec:level12}Data availability}

 The annotated dataset used for supervised evaluation is subject to preparations for manuscript publication. The identifiers of publications in the large 194,322 article dataset are available \cite{data_identifiers}.

\section{\label{sec:level13}Code availability}
All codes used in this study are publicly available \cite{article_code}.

\section{\label{sec:level14}Author contributions}
M.T., F.G., and S.P. conceived the original plan for the research and supervised the work. M.S. performed all computational and analysis work and drafted the manuscript. F.M. advised on the NLP training and application. Everyone contributed to data analysis and writing the manuscript.\\

\section{\label{sec:level15}Competing interests}
The authors declare no competing interests. \\

\begin{acknowledgments}
The authors thank the CSC-IT Center for Science in Finland for high performance computing resources. We acknowledge Mahboubeh Hadadian, Aleksi Kamppinen, Christer Söderholm and Ransell D'Souza for participating in evaluation dataset annotation and Emil Nuutinen for QA training scripts. Research was funded by the Research Council of Finland through grant number 345698. 
\end{acknowledgments}

\bibliographystyle{naturemag}
\bibliography{article}

\end{document}


\title{Supplementary Information \\ for \\ Question Answering models for information extraction \\ from perovskite materials science literature}

\author{M. Sipilä$^{1}$, F. Mehryary$^{2}$, S. Pyysalo$^{2}$, F. Ginter$^{2}$ and M. Todorović$^{1}$}
\affiliation{$^{1}$Department of Mechanical and Materials Engineering, University of Turku, Finland \\ 
$^{2}$Department of Computing, University of Turku, Finland}





\maketitle
\onecolumngrid

\section{\label{sec:level1}Data format conversion and duplicate removal}
ChemDataExtractor2 (CDE2) ready-made tools were used to convert the article files from Elsevier, arXiV and Royal Society of Chemistry to plain text. The format of publications downloaded through Springer Nature API was JATS-XML. We wrote a conversion script which removed all the XML-tags indicating figure, reference, mathematical equations and other information which could not be used by language models. Publications retrieved through Core API were in json-format. They were converted to plain text with the script we wrote, where abstract and fulltext were extracted from the json-file via json-tags 'abstract' and 'fullText'. Only the publications which contained at least the 'fullText'-tag were saved. Small cleaning script was also applied to Core articles, where possible italic and sub- or superscript tags were removed, if found. After converting the articles to plain text, some files were found to be empty, which is due to CDE2 conversion tools being uncapable to convert some aXiV articles. 
The empty files were removed from the dataset (see Table \ref{tab:conversion_statistics}). Because some conversion methods left unnecessary linebreaks in the text, plain text linebreaks were additionally cleared from all files.

\begin{table*}[h]
\caption{\label{tab:conversion_statistics} Article count after different stages of text processing and duplicate removal.}
\begin{ruledtabular}
\begin{tabular}{lllll}
 Data provider&After converting&After removing&After removing duplicates
&After removing duplicates\\ 
 &to plain text&empty articles& using DOI
&using vector representations\\
\hline
 Elsevier Article Retrieval API&113,438&113,438&113,438&113,438 \\
 Springer Nature API&34,395&34,395&34,379&34,379\\
 Core API&27,697&27,696&27,509&27,500\\
 arXiV API&3,797&3,709&2,407&2,371\\
 Royal Society of Chemistry&16,634&16,634&16,634&16,634\\
\end{tabular}
\end{ruledtabular}
\end{table*}

Next we ensured that our dataset consisted of only unique publications. We carried out duplicate removal by comparing the digital object identifiers (DOIs) of all the publications against the entire dataset. Because different data providers had different ways of encoding upper- and lowercase letters with DOIs, we converted all the DOIs to lowercase-only letters. We discovered 489 publications with DOI-based duplicates. When removing articles, we had to determine which data sources to prioritise for reliability. Our preferred removal order was: 1. arXiV, 2. Core, 3. Royal Society of Chemistry, 4. Springer and 5. Elsevier. The preference to remove first arXiV texts was based on the conversion from PDF to plain text, which is difficult and may lead to data loss. Core API publications did not need converting, but they were from multiple data sources for which the quality was not confirmed. Royal Society of Chemistry publications were in HTML-format, which can be slightly more difficult to convert to plain text than XML-formatted publication of Elsevier and Springer. The number of texts left after removing duplicates based on DOI is presented in Table~\ref{tab:conversion_statistics}.

Some of the publications acquired through arXiV API and Core API lacked a DOI. We inspected the uniqueness of these articles by converting them to token vectors with \textit{sklearn} CountVectorizer tool and calculating the cosine distance between the resulting vectors. If the similarity score was over 97 \%, the texts were found to be nearly identical and the duplicate was removed. With this approach, we excluded 36 arXiV articles and 9 Core articles.

\section{\label{sec:level1} Text snippet formatting}
Extracting information from full text publications with the QA method is inefficient, since the desired information (material-property relationship) is usually inside a couple of sentences. Using Elasticsearch search engine, we queried the articles by keywords and retained them for further use. Here, we generated a list of synonyms for material names and bandgap (see Table \ref{tab:synonymes}).
Both the material and property search was case-sensitive because the chemical element names are case sensitive; although bandgap is not, this approach is general and valid for future applications.

\begin{table*}[t]
\caption{\label{tab:synonymes}%
Different synonyms for bandgap, methylammonium lead iodide, methylammonium lead bromide, formamidinium lead iodide, cesium lead tri-iodide and cesium lead tri-bromide.
}
\begin{ruledtabular}
\begin{tabular}{ccc}
\textrm{Synonyms for Bandgap}&
\textrm{Synonyms for MAPI}&
\textrm{Synonyms for CsPbBr3}\\
\colrule
bandgap & methylammonium lead iodide & cesium lead bromide\\
band-gap & methylammonium lead triiodide & Cesium lead bromide\\
bandg gap & CH3NH3PbI3 & cesium lead tribromide\\
bandedge & MAPbI3 & Cesium lead tribromide\\
band-edge &(CH3NH3)PbI3 &CsPbBr3\\
band edge &\verb|[|CH3NH3PbI3\verb|]| & CsBr3Pb \\
Bandgap & lead methylammonium tri-iodide & Cs(PbBr3)\\
Band-gap & methyl amine lead(II) iodide & Cs$[$PbBr3$]$\\
Band gap & MAPI & cesium leadbromide\\
Bandedge & (CH3 NH3)PbI3 & Cesium leadbromide\\
Band-edge & CH3 NH3PbI3 \\
Band edge & CH3 NH3 PbI3 \\
 &methylammonium leadiodide & \\
 &methyl amine lead iodide & \\
 &Methylammonium lead iodide & \\
 &Lead methylammonium tri-iodide & \\
 &Methyl amine lead iodide & \\
 &Methananium lead(II) iodide & \\
 &Methylammonium leadiodide & \\
 &Methylamine lead iodide & \\
\hline
\textrm{Synonyms for MAPB}&
\textrm{Synonyms for FAPI}&
\textrm{Synonyms for CsPbI3}\\
\colrule
methylammonium lead bromide & formamidinium lead iodide & cesium lead iodide\\
Methylammonum lead bromide & Formamidinium lead iodide & Cesium lead iodide\\
methylammonium lead tribromide & formamidinium lead triiodide & cesium lead triiodide\\
Methylammonium lead tribromide & Formamidinium lead triiodide & Cesium lead triiodide\\ 
CH3NH3PbBr3 & CH(NH2)2PbI3 & CsPbI3\\
MAPbBr3 & FAPbI3 & CsI3Pb\\
(CH3NH3)PbBr3 & (CH(NH2)2)PbI3 & Cs(PbI3)\\
$[$CH3NH3$]$PbBr3 & lead formamidinium tri-iodide & Cs$[$PbI3$]$\\
lead methylammonium tri-bromide & Lead formamidinium tri-iodide & cesium lead tri-iodide\\
Lead methylammonium tri-bromide & formamidinium lead(II) iodide & Cesium lead tri-iodide\\
methyl amine lead bromide & Formamidinium lead(II) iodide & cesium leadiodide\\
Methyl amine lead bromide & FAPI  &Cesium leadiodide\\
methylamine lead bromide & (CH (NH2)2)PbI3\\
Methylamine lead bromide & CH NH2 2PbI3\\
MAPB & CH NH22PbI3\\
(CH3 NH3)PbBr3 & formamidinium leadiodide\\
$[$CH3 NH3$]$PbBr3 & Formamidinium leadiodide\\
CH3 NH3PbBr3 & HC(NH2)2PbI3\\
CH3 NH3 PbBr3 & (HC(NH2)2)PbI3\\
methylammonium leadbromide & (HC (NH2)2)PbI3\\
Methylammonium leadbromide & HC NH2 2PbI3\\
methylamine lead bromide & HC NH22PbI3\\
Methylamine lead bromide & \\
CH3NH3PbBr3
\end{tabular}
\end{ruledtabular}
\end{table*}

Next, we parsed the full text for seven-sentence snippets. Snippet length was chosen by distance analysis below. In Figure. \ref{fig:distances}, the snippets are grouped by how far away the material name, property name and the unit name are from each other in sentences. The distance is computed so that if the material name ('MAPI'), property name ('bandgap') and unit name ('eV') are found within one sentence, the distance is 0. If they are spread out between two sentences, the distance is 1 and so on. This analysis includes snippets for all five perovskite materials MAPI, FAPI, MAPB, CsPbI3 and CsPbBr3, with the total quantity of snippets being 13,943. Table \ref{tab:distances} encodes this information as proportions. In most snippets, all the information can be found inside one sentence, but this still constitutes only 40.4 \% of the total dataset. However, the amount of information in sentences further away from each others decreases fast. We estimated that longer than seven sentence snippets would not lead to extracting substantially more values, but could possibly lead to extracting more erroneous information.

\begin{minipage}{\textwidth}
  \begin{minipage}[b]{0.49\textwidth}
    \centering
\includegraphics[width=0.7\textwidth]{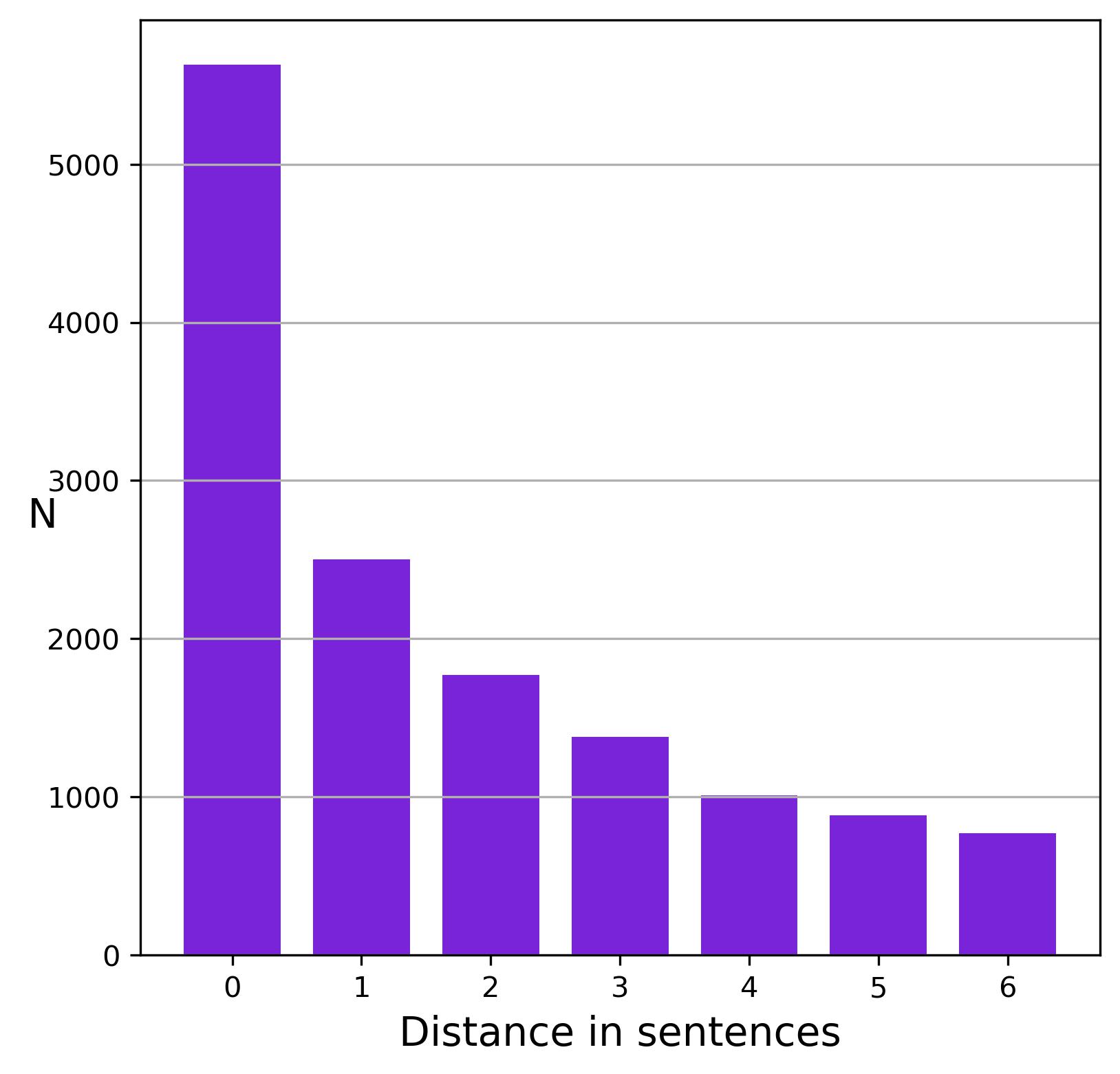}%
    \captionof{figure}{\label{fig:distances} Snippet count with varying sentence distances between material, property and the unit names.}
  \end{minipage}
  \hfill
  \begin{minipage}[b]{0.49\textwidth}
\centering
\begin{tabular}{cc}
\hline
\hline
\textrm{Distance}&
\textrm{Percentage}\\
\colrule
0 & 40.4 \%\\
1 & 17.9 \%\\
2 & 12.7 \%\\
3 & 9.9 \% \\
4 & 7.3 \%\\
5 & 6.3 \%\\
6 & 5.5 \%\\
\hline
\hline
\end{tabular}
      \captionof{table}{\label{tab:distances} Proportion of snippets with different sentence distances separating material, property and unit names.}
    \end{minipage}
  \end{minipage}
\\
\\

To ensure the seven-sentence format of snippets, we used the sentence splitter tool from the python library \textit{nltk.data} \cite{nltk}. The seven-sentence window was iterated through the text, and when the window captured the material name, property name (see Table \ref{tab:synonymes} for synonyms) and the unit 'eV', the process was paused. The snippet window was shifted three sentences forward to prevent any information loss due to valuable information located in the last sentences. We tested different sentence shifts: with three sentences, the mentions of material name, property name and the unit were best centred in the snippet. 
The selected text segments were assigned a unique and identifiable name, which consisted of the DOI (or \texttt{arXiV id} or \texttt{Core id}) and index of the character in the snippet which had matched the regular expression of material and property.\\
\\
The BERT models can process texts with the maximum token number of 512 without any information loss. To investigate whether the 7-sentence snippets cross this limit, we computed the percentage of 7,283 MAPI and 1,251 FAPI snippets with token count below 512. The data in Table \ref{tab:token_lengths} illustrates that the fraction of snippets which exceed this limit is small ($\sim$ 10 \%). In the cases when the context document exceeds the 512 token BERT limit, the text is truncated from the end of the snippet. Since our procedure was designed to center the valuable information in the snippet, the truncation should not affect relevant information, and information loss should be minimal. Any truncation could also be avoided using shorter snippets (6 sentences or below) but this could then lead to loss in context.

\begin{table*}[h]
\centering
\caption{\label{tab:token_lengths}
Percentage of snippets extracted for MAPI and FAPI which boast a token number below 512.}
\begin{ruledtabular}
\begin{tabular}{lccc}
\textrm{Model}& \textrm{MAPI}& \textrm{FAPI}\\ \hline
BERT & 90.1 \%  & 85.1 \% \\
SciBERT & 92.8 \% & 89.1 \%\\
MatBERT & 94.3 \% & 90.6 \%\\
MaterialsBERT & 93.0 \% & 89.0 \%\\
MatSciBERT & 92.8 \% & 89.1 \%\\
\end{tabular}
\end{ruledtabular}
\end{table*}
\vspace{1cm}

\section{\label{sec:level1}Training the machine learning models}

We trained the BERT models to the QA task using 2 epochs and batch size of 12, because these lead to the best results in preliminary tests with the QA training. The learning rate was optimised for each BERT separately between values 1e-5, 3e-5, 5e-5 and 8e-5. Training loss was used as the defining metric. Table \ref{tab:losses} summarises the training losses of models and the learning rates tested. The best learning rate was 3e-5 for MatBERT, MatSciBERT, SciBERT and BERT and 5e-5 for MaterialsBERT. The best training loss for each model is marked in bold in Table \ref{tab:losses}.

We trained the models with these selected learning rates using seeds 12, 16, 30, 36 and 42. The seed variation was done to ensure reliable comparison between models, because different seeds could have a different impact to the trained model. The models were trained on NVIDIA Volta V100 GPU. The average run times and standard deviations of run times for training models across the seeds are collected to Table \ref{tab:losses}.

\begin{table*}[h]
\caption{\label{tab:losses}%
Training loss of different materials science BERTs when trained with different learning rates (lr)}.
\begin{ruledtabular}
\begin{tabular}{lccccc}
\textrm{}&
\textrm{Training loss}&\textrm{Training loss}&\textrm{Training loss}&\textrm{Training loss}&\textrm{Average training}\\
\textrm{Language model}&
\textrm{lr=1e-5}&
\textrm{lr=3e-5}&
\textrm{lr=5e-5}&
\textrm{lr=8e-5}&
\textrm{time (min)}\\
\colrule
MatBERT & 1.190 & \textbf{1.073} & 1.096 & 1.230 & 135 (±19)\\
MatSciBERT & 1.218 & \textbf{1.041} & 1.053 & 1.142 & 128 (±8)\\
MaterialsBERT & 1.792 & 1.549 & \textbf{1.515} & 1.559 & 121 (±40)\\
SciBERT & 1.125 & \textbf{0.978} & 0.999 & 1.096 & 140 (±16)\\
BERT & 1.165 & \textbf{0.960} & 0.964 & 1.007 & 123 (±12)\\
\end{tabular}
\end{ruledtabular}
\end{table*}

\section{\label{sec:level1}Postprocessing of QA-returned answers}
In information extraction (IE) tasks, the extracted information is postprocessed (sometimes extensively) to ensure the good quality of the answers and to make it possible to visualise the results. Our aim was to define a postprocessing scheme which would not change the answers much, but would remove technically correct answers that have low information content (not useful). To address the question of how much values should be postprocessed, we defined two different postprocessing schemes to be tested with the evaluation dataset:

\begin{enumerate}
\item  \textit{Partial postprocessing (PP)}. Here the returned QA answers are filtered so that all the answers that do not contain any numbers are omitted. For example, this excludes the QA returned answer 'narrow' in response to the question 'What is the numerical bandgap of this material?'. From the answers where a number was present, we removed extra letters. In this way the answer '1.5 eV' would be processed to '1.5', in order to calculate the exact match with the annotated value.
\item \textit{Full postprocessing (FP)}. Here all the answers which do not contain any numbers are discarded, and also the answers which contain some other letters than following: 'a, n, d, t, o, e, v'. These characters were selected since they are the ones that also have been annotated in the annotation dataset ('and', 'to' and 'eV'). Answers where the first character was dash were also removed, since we could not be sure whether they were negative or just a part of range in the original snippet. Finally, any additional letters were removed similarly as in the PP.
\end{enumerate}

Table \ref{tab:MatBERT} illustrates the effect of the different postprocessing schemes on the answers returned by the QA MatBERT with different confidence thresholds. The testing was performed on the 4-fold cross-validation (CV) validation dataset and we computed the precision, recall and F1-score by averaging from 5 statistically different models. 

As more post-processing was introduced, the precision and the recall both improved for all the thresholds. At threshold 0.2 (best result), F1-score improved from 58.38 to 60.39. We observed the same effect with the other language models tested (BERT, SciBERT, MatSciBERT and MaterialsBERT), so we adopted full postprocessing in all further analysis. We also found that the CDE2 extraction method also has an extensive postprocessing scheme, which indicates that this is the recommended practice.

\begin{table*}[h]
\caption{\label{tab:MatBERT}%
Evaluation results for MatBERT with different postprocessing schemes of CV validation set: PP for partial postprocessing and FP for full postprocessing. As evaluation metrics we use precision (P), recall (R) and F1-score (F1). The best results for each model are indicated in bold. The values are presented as mean and (±) standard deviation in parentheses computed from 5 language models trained with different random seeds.
}
\begin{ruledtabular}
\begin{tabular}{llcccccccc}
\textrm{}&
\textrm{}&
\textrm{}&
\textrm{}&
\textrm{Threshold}&
\textrm{}&
\textrm{}&
\textrm{}&
\textrm{}&\\
\textrm{}&
\textrm{}&
\textrm{0.0124}&
\textrm{0.025}&
\textrm{0.05}&
\textrm{0.1}&
\textrm{0.2}&
\textrm{0.4}&
\textrm{0.8}&\\
\colrule

PP   & P & 37.23 (± 1.53) & 40.47 (± 1.62) & 44.78 (± 2.49) & 49.43 (± 3.24) & 54.12 (± 3.81) & 63.73 (± 6.17) & 80.20 (± 6.85)\\
                    & R & 75.50 (± 1.84) & 73.58 (± 1.68) & 70.99 (± 1.80) & 68.02 (± 1.84) & 63.61 (± 1.79) & 54.52 (± 2.98) & 26.48 (± 6.57)\\
                  & F1 & 49.84 (± 1.48) & 52.19 (± 1.34) & 54.88 (± 1.97) & 57.18 (± 2.19) & 58.38 (± 2.03) & 58.53 (± 2.60) & 39.34 (± 7.70)\\
                    \hline 
FP   & P & 39.29 (± 1.62) & 43.31 (± 1.87) & 47.99 (± 2.83) & 52.77 (± 3.70) & 57.33 (± 4.50) & 67.29 (± 6.79) & 85.28 (± 6.71) \\
                   & R & 77.21 (± 1.70) & 75.40 (± 1.72) & 72.23 (± 1.95) & 69.07 (± 2.17) & 64.09 (± 1.94) & 54.52 (± 2.98) & 26.48 (± 6.56)\\
                   & F1 & 52.05 (± 1.36) & 54.97 (± 1.40) & 57.61 (± 2.00) & 59.72 (± 2.21) & \textbf{60.39 (± 2.29)} & 59.98 (± 2.74) & 39.87 (± 7.70) \\   
\end{tabular}
\end{ruledtabular}
\end{table*}


Before plotting the values into histogram, we also removed the values which did not belong to the interquartile range between 10 and 90 of the values. This ensures that the quality of the extracted values stays high and omits clear outliers. Also, before visualising the results, the average values of the answers denoting range was taken (for example "1.5-1.6 eV" $\rightarrow$ "1.55 eV"). This approach ensured that when visualising the results, the emphasis was not in the range end points. The answers denoting two values were splitted to two values (for example "1.3 eV and 1.34 eV" $\rightarrow$ "1.3 eV" and "1.34 eV") to visualise them as separate datapoints. In visualising the results, the answers were also rounded to two decimals.
\pagebreak

\section{\label{sec:level1} Evaluation dataset}

In order to evaluate QA model performance, we generated a manually annotated dataset of 600 snippets. Manual annotation is a standard process in the NLP field to create a gold standard dataset by marking the correct labels or answers to the text manually by human experts. The snippets were from open access articles, to ensure that the entire evaluation dataset could be published for future benchmarking of IE tools in the field of materials science. There were 6 materials science expert annotators who took part in the annotation process, each annotating 200 snippets. Every snippet was annotated by two different people and contentious annotations were afterwards resolved in a meeting between two annotators and the annotation supervisor. 

The dataset of 600 seven-sentence snippets contained equal numbers of snippets based on five perovskite materials: methylammonium lead iodide (MAPI), methylammonium lead bromide (MAPB), formamidinium lead iodide (FAPI), cesium lead iodide (CsPbI$_{3}$) and cesium lead bromide (CsPbBr$_{3}$). Snippets were selected equitably from different data provider sources, since the quality of the text varies between publications from different data providers. Still, restrictions in open access article amounts limited the amount of snippets from certain data providers. Table \ref{tab:evaluation_dataset} illustrates the breakdown of snippets by data provider and material in the annotated dataset.

\begin{table*}[h]
\caption{\label{tab:evaluation_dataset}%
Counts of snippets in the evaluation dataset based on the data provider and the material.
}
\begin{ruledtabular}
\begin{tabular}{lcccccc}
\textrm{Material/Data provider}&
\textrm{arXiV}&
\textrm{Core}&
\textrm{Elsevier}&
\textrm{Springer}&
\textrm{RSC}&\\

\colrule
MAPI & 24 & 24 & 24 & 24 & 24\\
MAPB & 27 & 28 & 19 & 19 & 27\\
FAPI & 25 & 26 & 19 & 24 & 26\\
CsPbI3 & 26 & 27 & 27 & 14 & 26\\
CsPbBr3 & 24 & 24 & 24 & 24 & 24\\
\end{tabular}
\end{ruledtabular}
\end{table*}

The annotated bangaps were material specific: any bandgap values that annotators could relate to a different material were ignored. If the bandgap was in the format '1.5-1.6 eV' or '1.5 to 1.6 eV', the entire text span was accepted as correct since the range can be understood as one entity. Bandgap values which were separated by a comma ('1.5, 1.6 eV') or 'and' ('1.5 and 1.6 eV') were annotated as two separate values. Also, only the bandgaps which were reported in electronvolts were annotated.

\section{\label{sec:level1} Model performance on the evaluation dataset}

We evaluated all the QA models with the manually annotated evaluation dataset. The snippets in the evaluation dataset were input to the language models and we posed the question 'What is the numerical value of bandgap of [material]?' for 5 different perovskite materials. We modified the QA models to always return the 6 top answers for each snippet, i.e. the answers the model had assigned with the highest confidence scores. As an output, the model returns a table of 6 most likely answers for each snippet, combined with their confidence scores and text span starting and ending indices. We collected the snippets of all of the models trained with different seeds, divided them into 4 folds and performed 4-fold cross-validation (CV) and then compared the output answers to the golden standard answers with the evaluation script. The mean and standard deviation of the evaluation metrics were calculated from models trained with different seeds. 

Table \ref{tab:full_postprocessing_validation} contains the CV validation set results for all language models, fully postprocessed. The best F1-scores are marked in bold and indicate that thresholds 0.1 or 0.2 are optimal for all other models except BERT (the optimal threshold for BERT is 0.025).

\begin{table*}[ht]
\centering
\caption{\label{tab:full_postprocessing_validation}
Validation dataset results with full postprocessing for BERT, SciBERT, MatBERT, MaterialsBERT and MatSciBERT with different thresholds. The best results for each model are indicated in bold. Evaluation metrics are precision (P), recall (R) and F1-score (F1).  The values are presented as mean and (±) standard deviation in parentheses computed from 5 language models trained with different random seeds.
}
\begin{ruledtabular}
\begin{tabular}{llcccccccc}
\textrm{}&
\textrm{}&
\textrm{}&
\textrm{}&
\textrm{Threshold}&
\textrm{}&
\textrm{}&
\textrm{}&
\textrm{}&\\
\textrm{}&
\textrm{}&
\textrm{0.0125}&
\textrm{0.025}&
\textrm{0.05}&
\textrm{0.1}&
\textrm{0.2}&
\textrm{0.4}&
\textrm{0.8}&\\
\colrule

BERT  & P & 41.65 (± 4.76) & 44.83 (± 5.87) & 48.05 (± 6.02) & 54.23 (± 7.13) & 60.63 (± 7.63) & 67.66 (± 6.61) & 81.97 (± 7.68) &\\
                    & R & 59.87 (± 5.00) & 54.99 (± 5.26) & 49.44 (± 5.75) & 43.51 (± 5.17) & 36.63 (± 4.89) & 26.89 (± 4.19) & 11.68 (± 2.72) &\\
                    & F1 & 48.75 (± 2.42) & \textbf{48.88 (± 2.67)} & 48.20 (± 2.58) & 47.67 (± 2.11) & 45.15 (± 3.03) & 38.09 (± 3.52) & 20.31 (± 4.17) & \\ 
                    \hline 
SciBERT & P & 38.51 (± 1.88) & 42.09 (± 1.56) & 46.19 (± 1.74) & 49.97 (± 2.22) & 54.96 (± 2.31) & 62.73 (± 3.09) & 79.66 (± 3.72) &\\
                    & R & 77.86 (± 3.29) & 74.98 (± 2.77) & 71.74 (± 2.41) & 67.25 (± 2.52) & 62.18 (± 2.48) & 50.53 (± 3.27) & 22.29 (± 3.52) &\\
                    & F1 & 51.50 (± 1.94) & 53.88 (± 1.49) & 56.17 (± 1.46) & 57.29 (± 1.69) & \textbf{58.29 (± 1.40)} & 55.84 (± 1.78) & 34.65 (± 4.03) & \\
                    \hline 

Mat- & P & 39.29 (± 1.62) & 43.31 (± 1.87) & 47.99 (± 2.83) & 52.77 (± 3.70) & 57.33 (± 4.50) & 67.29 (± 6.79) & 85.28 (± 6.71) & \\
BERT                  & R & 77.21 (± 1.70) & 75.40 (± 1.72) & 72.23 (± 1.95) & 69.07 (± 2.17) & 64.09 (± 1.94) & 54.52 (± 2.98) & 26.48 (± 6.56) & \\
                    & F1 & 52.05 (± 1.36) & 54.97 (± 1.40) & 57.61 (± 2.00) & 59.72 (± 2.21) & \textbf{60.39 (± 2.29)} & 59.98 (± 2.74)& 39.87 (± 7.70) & \\
                    \hline
                    
Materials-  & P & 41.65 (± 2.43) & 45.10 (± 3.61) & 49.80 (± 3.39) & 54.65 (± 3.68) & 60.22 (± 3.75) & 71.10 (± 5.54) & 90.54 (± 6.21) &      \\
BERT                  & R&  75.18 (± 3.82) & 70.97 (± 5.26) & 65.62 (± 6.33) &61.12 (± 6.76) & 53.96 (± 8.77) &43.64 (± 10.06) & 17.53 (± 6.47) &       \\
                    & F1& 53.50 (± 1.77) & 54.94 (± 2.53) & 56.35 (± 2.59) &\textbf{57.33 (± 2.63)} & 56.40 (± 5.02) & 52.98 (± 7.73) & 28.75 (± 9.49) &       \\
                    \hline
MatSci- & P& 47.40 (± 5.21) & 51.59 (± 6.26) &58.49 (± 6.32) & 65.22 (± 6.50) & 69.49 (± 3.50) & 75.63 (± 5.44) & 89.94 (± 8.54) &\\
  BERT      & R & 73.46 (± 6.13) & 70.70 (± 7.26) & 67.83 (± 8.90) & 64.40 (± 9.11) & 56.27 (± 11.13) &43.82 (± 11.66) & 15.30 (± 7.66) & \\
                    & F1 & 57.16 (± 2.17) & 58.98 (± 1.89) & 62.01 (± 2.00) & \textbf{63.93 (± 2.14)} & 61.30 (± 6.03) & 54.16 (± 8.83) & 25.07 (± 11.27) &\\
\hline
\end{tabular}
\end{ruledtabular}
\end{table*}
\vspace{1cm}

In typical applications of QA models only the top  (most confident) answer is returned, so we computed the corresponding metrics for this use case and collected them to Table \ref{tab:first_answer}. As observed from Table \ref{tab:first_answer}, the F1-scores for selecting the top answer and ones based on the threshold 0.1 or 0.2 are very similar. 
To explain this, we calculated the model confidences associated with each of the 6 top answers (see Table \ref{tab:average_confidence_score}). From Table \ref{tab:average_confidence_score}, we note that after the first answer, the average confidence score tends to decrease from 0.6-0.7 by an order of magnitude. Consequently the confidence threshold of 0.1 or 0.2 in many cases simply selects the first answer.

\begin{table*}[h]
\centering
\caption{\label{tab:first_answer}
Results when only the first answer is selected, calculated with the evaluation dataset. Evaluation metrics are presented as F1-score (F1), precision (P) and recall (R). Standard deviation is obtained based on the answers from the models trained with 5 different seeds.}
\begin{ruledtabular}
\begin{tabular}{lcccc}
\textrm{Model}& \textrm{F1}& \textrm{P}& \textrm{R}\\ \hline
BERT & 44.52 (± 2.62) & 59.42 (± 7.38) & 36.46 (± 6.22)  \\
SciBERT & 57.93 (± 1.13) & 54.32 (± 1.98) & 62.2 (± 2.36) \\
MatBERT& 60.43 (± 1.99) & 56.54 (± 4.59) & 65.17 (± 1.41) \\
MaterialsBERT& 57.55 (± 5.17) & 61.37 (± 4.26) & 55.21 (± 9.56) \\
MatSciBERT& 62.81 (± 5.98) & 70.02 (± 4.77) & 58.47 (± 11.98) \\
\hline
\end{tabular}
\end{ruledtabular}
\end{table*}
\vspace{1cm}

\begin{table*}[h]
\caption{\label{tab:average_confidence_score}%
Average confidence score with standard deviation of 6 first answers for different models.
}
\begin{ruledtabular}
\begin{tabular}{lcccccccc}

\textrm{Model}&
\textrm{First answer}&
\textrm{Second answer}&
\textrm{Third answer}&
\textrm{Fourth answer}&
\textrm{Fifth answer}&
\textrm{Sixth answer}&\\
\colrule
BERT & 0.719 (±0.268) & 0.044 (±0.085) & 0.013 (±0.026) & 0.006 (±0.016) & 0.003 (±0.007) & 0.002 (±0.005)\\
SciBERT & 0.676 (±0.268) & 0.063 (±0.098) & 0.019 (±0.032) & 0.009 (±0.019) & 0.005 (±0.011) & 0.003 (±0.007) \\
MatBERT & 0.668 (±0.263) & 0.065 (±0.097) & 0.020 (±0.034) & 0.009 (±0.019) & 0.005 (±0.009) & 0.003 (±0.006)\\
MaterialsBERT & 0.645 (±0.270) & 0.058 (±0.090) & 0.018 (±0.034) & 0.009 (±0.022) & 0.005 (±0.010) & 0.003 (±0.007)\\
MatSciBERT & 0.710 (±0.271) & 0.052 (±0.093) & 0.016 (±0.032) & 0.007 (±0.017) & 0.003 (±0.009) & 0.002 (±0.006)\\
\end{tabular}
\end{ruledtabular}
\end{table*}

\pagebreak
\section{\label{sec:level1} The performance of generative models}

We compared the performance of QA models also to the IE capabilities of four generative models. To ensure that the comparison from the user’s perspective was as similar as possible to the QA method, we opted for zero-shot IE, although better results might have been achieved with multiple consecutive prompts. We tested four different prompts, which range from short and general to long and detailed. The aim was to test how the results change when we add more definitions to the prompt. For each model, we selected the one which produced the best results. In the following prompts, [material] was replaced with the material of interest:\\

\begin{enumerate}
\item "What is the numerical value of bandgap of [material] in the following context? Context:"+text
\item "What is the numerical value of bandgap, or values of bandgaps, of pure [material] in the following context? Context:"+text
\item "In the following context, what is the numerical bandgap value or bandgap values of pure [material]? Only answer a single number in eV units or, in case of multiple reported bandgaps, numbers in eV units. Answer 'none' if there is no statement about the bandgap, or if there is no statement of numerical bandgap. Context: "+text
\item "In the following context, what is the absolute bandgap value or bandgap values of pure [material]? Only answer a single number in eV units or, in case of multiple reported bandgaps, numbers in eV units. Answer 'none' if there is no statement about the bandgap, or if there is no statement of absolute bandgap. Do not report bandgaps of doped materials, or if the text does not indicate to which material the bandgap belongs. Context: "+text
\end{enumerate}

To ensure a reliable comparison of the results produced by the generative models, it was necessary to postprocess the outputs. Answers generated by different models were different, so we created one postprocessing scheme for Mixtral-8 and Llama3 and a different one for GPT models (two in total). The main difference between the answers generated by Mixtral-8, Llama3 and GPT models was the formatting of the answers: Mixtral-8 and Llama3 returned complete sentences with much explanatory text. GPT model answers had more consistent format throughout the answers and less explanatory text, with multiple values usually separated by commas. The following steps were taken when postprocessing Mixtral-8 and Llama3 answers:

\begin{enumerate}
    \item  [1.]  Erase the material name from the answer.
    \item  [2.] Set the answer to empty if the word 'None' or similar was present in the answer.
    \item [3.] Omit the answers if the letters 'eV' were not present.
    \item [4.] Select all the substrings from the answer which contain numbers, commas, dots, dashes, the word 'to' and a whitespace after them and erase everything else.
    \item [5.] For each substring,  select the first and last occurrence of a number and remove all characters before and after them. Check if the substring can be found from the snippet and if so, assign the index denoting the beginning and ending of the answer.
    \item [6.] If the substring did not match any string in the snippet, compare individual floats (for ex. '1.5' and '1.6' in the substring '1.5 eV, 1.6')in the substring to the snippet and assign beginning and ending indices if found.
    \item [7.] If the answer can not still be found from the snippet, it is marked as hallucinated.
\end{enumerate}

The postprocessing script for GPT-3.5 and GPT-4 was similar to the one with Mixtral and Llama, except in one stage: in the point four we did not select all the substrings, but rather selected the single values to process. In the case of the multiple answers, we split the answer by the comma or the word 'and' and selected the separated values. The evaluation results of generative models after postprocessing for different prompts are presented in Table \ref{tab:gen_results}. The temperature of the models was set to 0, and the seed in GPT-3.5 and GPT-4 was set to 12. Based on the Table \ref{tab:gen_results} we selected the suitable prompts (the prompt exhibiting the highest F1-score) for each model to be used in the manuscript.

\begin{table*}[h]
\centering

\caption{\label{tab:gen_results}
Generative model results for different prompts. The highest F1-score values are marked with bold.}

\begin{ruledtabular}
\begin{tabular}{lcccccccccccccccccc}
\textrm{}&\textrm{Prompt1} &\textrm{}&\textrm{}&\textrm{}&\textrm{Prompt2} &\textrm{}&\textrm{}&\textrm{}&\textrm{Prompt 3} &\textrm{}&\textrm{}&\textrm{}&\textrm{Prompt 4}\\ 

\textrm{}&\textrm{F1}&\textrm{P}&\textrm{R}&\textrm{H}&\textrm{F1}&\textrm{P}&\textrm{R}&\textrm{H}&\textrm{F1}&\textrm{P}&\textrm{R}&\textrm{H}&\textrm{F1}&\textrm{P}&\textrm{R}&\textrm{H}
\\ \hline

Mixtral-8 & 51.40 & 39.20 & 74.64 & 34 & \textbf{54.08}& 43.57& 71.29& 39& 52.27& 38.84& 79.90& 42& 51.34& 38.26& 77.99& 51\\

Llama3 & 38.64& 26.24& 73.21& 119& 38.91& 26.25& 75.12& 110& \textbf{42.18}& 30.49& 68.42& 161& 32.86& 28.52& 38.76& 98\\

GPT-3.5& 52.12 & 39.51 & 76.56& 16& 53.28& 39.95&79.90& 43& 53.63& 41.41& 76.08& 11& \textbf{54.58}& 42.7& 75.60& 14\\

GPT-4 & \textbf{69.26} & 60.57 & 80.86& 1& 65.68& 53.45& 84.17 & 6& 62.14& 49.57& 83.25& 2&65.65& 54.4& 82.78& 2\\

\hline
\end{tabular}
\end{ruledtabular}
\end{table*}
\vspace{1cm}

\section{\label{sec:level1}Extracting MAPI and FAPI values with all models}

Extracted bandgap values for MAPI and FAPI were postprocessed and encoded in Table \ref{tab:full_extraction_statistics}. The average execution time of the QA models when extracting MAPI values on one GPU was 164 min and for FAPI the average execution time was 30 min. The execution time of CDE2 when extracting MAPI values on one GPU was 235 min and for FAPI 40 min. Table \ref{tab:boxplot_characteristics} contains the statistical characteristics of Figure 7 in the main article.

\begin{table*}[h]
\centering
\caption{\label{tab:full_extraction_statistics}
Statistical characteristics of N bandgap values for  MAPI and FAPI extracted with five different language models. The range is defined as the energy difference between maximum and minimum extracted bandgap values. The SD denotes the standard deviation of the extracted bandgap values.}
\begin{ruledtabular}
\begin{tabular}{lcccccccccccc}
\textrm{}&\textrm{MAPI}&\textrm{}&\textrm{}&\textrm{}&\textrm{}&\textrm{}&\textrm{FAPI}&\textrm{}&\textrm{}&\textrm{}\\
\textrm{Model}&\textrm{N} &\textrm{Mean}&\textrm{Median}&\textrm{Mode}&\textrm{SD}&\textrm{Range}&\textrm{N}&\textrm{Mean}&\textrm{Median}&\textrm{Mode}&\textrm{SD}&\textrm{Range}\\ \hline

BERT & 3,995 & 1.66 & 1.57 & 1.55 & 0.279 & 1.70 & 993 & 1.55& 1.49& 1.48 & 0.152& 0.93\\

SciBERT & 2,017 & 1.59 & 1.56 & 1.55& 0.096 & 0.61 & 413 & 1.49 & 1.48 & 1.48 & 0.152 & 0.27\\

MatBERT & 2,058 & 1.57 & 1.56 & 1.55 & 0.054 & 0.28 & 397 & 1.49 & 1.48 & 1.48& 0.041& 0.22\\

MaterialsBERT & 2,852 & 1.61 & 1.57 & 1.55 & 0.161 & 0.98 & 517 & 1.49& 1.48 & 1.48& 0.054& 0.35\\

MatSciBERT & 2,733& 1.59 & 1.57 & 1.55 & 0.088 & 0.55 & 536 & 1.50 & 1.48& 1.48& 0.071& 0.59\\

CDE2 & 953 & 1.60 & 1.55 & 1.55 & 0.575 & 6.99 & 49 & 1.49 & 1.48& 1.48& 0.460& 3.86\\
\hline
\end{tabular}
\end{ruledtabular}
\end{table*}
\vspace{1cm}

\begin{table*}[h]
\caption{\label{tab:boxplot_characteristics}%
Statistical characteristics of the boxplot (Figure. 7) in the article, indicating the values for the minimun, maximum, median, lower quartile (LQ) and upper quartile (UQ) of the distributions. 
}
\begin{ruledtabular}
\begin{tabular}{lcccccccccc}
\textrm{Material} &\textrm{Min} &\textrm{Max} &\textrm{Med} &\textrm{LQ} &\textrm{UQ}\\
\colrule
MAPI & 1.45 & 1.69 & 1.56 & 1.54 &1.60\\
MAPB & 1.97 & 2.44 & 2.26 & 2.16 & 2.30\\
FAPI & 1.41 & 1.57 & 1.48 & 1.47 & 1.51\\
CsPbI3 & 1.45 & 2.17 & 1.73 & 1.72 & 2.17\\
CsPbBr3 & 2.07 & 2.55 & 2.30 & 2.25 & 2.37\\
\end{tabular}
\end{ruledtabular}
\end{table*}

[1] S. Bird, E. Loper, and E. Klein, Natural language processing with Python: Analyzing text with the natural language toolkit.
(2019).